\documentclass[onecolumn,showpacs,preprintnumbers,amsmath,amssymb]{revtex4}

\usepackage{graphicx}
\usepackage{dcolumn}
\usepackage{bm}

\def\be{\begin{equation}}
\def\ee{\end{equation}}

\begin{document}

\title{Quantum mechanics on non commutative spaces and squeezed states: a 
functional approach.}

\author{Musongela Lubo}
 \affiliation{The Abdus Salam International Centre for Theoretical Physics I.C.T.P\\
	P.O.Box
586\\
34100 Trieste, Italy.\\
}
\email{muso@ictp.trieste.it}

\date{\today}
\begin{abstract}
 We review here the quantum mechanics of some noncommutative theories in which
 no state saturates
simultaneously all the non trivial Heisenberg uncertainty relations.
We show how the difference of structure between  the
Poisson brackets  and the commutators in these  theories  generically leads to a 
harmonic oscillator whose positions and momenta mean values are not strictly equal to the 
ones predicted 
by classical mechanics. 

This raises the question of the nature of quasi classical states in these models.
We propose an extension based on a variational
principle. The action considered is the sum  of the absolute values of the 
expressions
associated to the non trivial Heisenberg  uncertainty relations. We first verify
that our proposal works in the usual theory i.e we recover the known Gaussian 
functions. Besides them, we find other states which can be expressed as products 
of Gaussians with specific  hyper geometrics.

We illustrate our construction in two models defined on a four dimensional
phase space: a model endowed with a minimal length uncertainty
 and the non commutative plane. Our proposal leads
to   second order partial differential equations. We find analytical solutions 
in specific cases. We briefly discuss how our proposal may be applied to the fuzzy sphere
and analyze its shortcomings.
\end{abstract}

\maketitle

\section{Introduction}
    Many works have been devoted to noncommutative quantum theories recently.
One of the main motivations  is the hope that a non trivial
structure of space time at small distances may give birth to theories with 
better ultraviolet behaviors. 

There are two ways of apprehending non commutative theories. The first one
postulates a modification of the commutation relations from the
start. This is the case for example of J.Madore \cite{madore}. The
second one, in the wake of E.Witten and N.Seiberg \cite{witten},
considers non commutative theories as  low energy phenomenological
implications  of string theory. The philosophy of this work  relies
on the first approach.

Many non commutative quantum theories have not  fulfilled the initial hope
concerning the convergence of Green functions \cite{phi4}. One of the remarkable
cases which do not fall in this category has been formulated by
Kempf-Mangano-Mann(KMM) \cite{kmm3,kmm4,kmm5}. This could be achieved thanks to a careful analysis
of the states physically allowed in the theory. With that idea in mind, it was
suggested that the analysis about the loss of causality may need a more detailed 
treatment \cite{lubo}. 

Some efforts  have been devoted to the understanding of quantum mechanics when
non commutativity sets in. For example, H.Falomir et al. \cite{falomir} have studied
the Bohm-Aharonov effect in this context, obtaining for the deformation
parameter new bounds which are consistent with previous ones \cite{oldbound}. The
noncommutative oscillator in arbitrary dimension has been analyzed by
A.Hatzinikitas et al \cite{hatzi} while R.Banarjee \cite{banar} has shown 
its link to dissipation. The relation with usual canonical
variables has been analyzed by A.smailagic and E.Spallucci \cite{smail} and a path integral
formulation proposed by C.Acatrinei \cite{aca}. Some theories which also have
special fundamental structures have been used in the study of  physical processes like
black hole evaporation \cite{brout,muso}.

Our aim is not to solve a specific problem in this framework but to try to
understand the structure of some meaningful sets of states. In a recent
work, K.Bolonek and P.Kosinski \cite{kosi} have shown that all the non trivial
Heisenberg uncertainties can not be satisfied simultaneously in a theory where
the  commutators of the positions  are non vanishing constants. This raises an
 important question
since in the unmodified theory, the states which saturate all the non trivial 
 Heisenberg
uncertainties are also the ones which lead to classical trajectories. In 
these states, the mean values of the position and momentum operators
reproduce the behavior of the classical Hamilton solution for the harmonic
oscillator, the relative uncertainties being negligible \cite{cohen}.

Adapting  for our
purposes an idea  recently used by
S.Detournay, Cl.Gabriel and Ph.Spindel  \cite{spindel}, we  construct 
a  functional which attains its minima on the coherent 
 states in the usual theory. We then generalize this functional for the non commutative
theories we are interested in.  The minima of these functionals are, in our opinion, 
candidates to the status of  squeezed states. Our work is somehow similar in spirit with
the paper by  G. Dourado Barbosa \cite{bre} but the approach and the tools are different.

This paper is organized as follows. 
The second section is devoted to the impossibility of
saturating the non trivial  Heisenberg uncertainties in two particular
noncommutative theories. In the third section we show that 
for a generic state of a harmonic oscillator on the non commutative 
plane or in the KMM model, the mean positions do not reproduce rigorously  the behavior obtained using
 a classical analysis; although  experimentally  unobservable, this fact raises an important 
 conceptual question. In the  fourth section, we show explicitly how our proposal 
works in the usual one dimensional case; we verify that the usual squeezed states satisfy
the second order differential equation which comes from our variational
principle.
In the fifth section we apply the same method to the two models studied in
section $2$. We obtain  that the states for which the functionals are extremal
verify  second order partial differential equations. We give some exact solutions
but it remains to be proved that they effectively realize the minimum of the functional.
In the sixth section we briefly discuss how our procedure can be extended to the fuzzy sphere.   
The seventh section is devoted to a discussion of our results.

\section{Examples of high dimensional 
theories with non trivial Heisenberg uncertainties that cannot be satisfied
simultaneously}

In  usual quantum theory, the coherent states satisfy {\em simultaneously}
the relations 
\be
\label{eq1}
\Delta x_i \Delta p_i = \frac{ \vert \langle [\hat x_i, \hat p_i] \rangle 
\vert}{2} = \frac{\hbar}{2} \ee
 for each index $i$. On the
contrary, a product such as $\Delta x_1 \Delta p_2$  can be arbitrary since the
commutator of the  associated variables vanishes. The first type of
uncertainty will be called {\em non trivial}. The aim of this section is to
show that in some  two dimensional models  all the non trivial uncertainties
cannot be saturated at the same time. The same reasoning works in higher
dimensions. 

To proceed, we will associate to each fixed state $\vert \psi \rangle$ an operator $\hat a_\psi$
whose action on an arbitrary state $\vert \phi \rangle$ is defined by
\be
\label{neweq2}
 \hat a_\psi \vert \phi \rangle = \left[ \hat x - \langle \hat x \rangle_\psi I + 
 \frac{\langle [\hat x, \hat p] \rangle_{\psi} }{2 (\Delta_{\psi} p)^2} 
  ( \hat p - \langle \hat p \rangle_\psi I)  \right] \vert \phi \rangle \quad ,
\ee
$I$ being the identity operator.
The states $\vert \psi \rangle $ which satisfy Eq.(\ref{eq1}) and thus saturate the 
Heisenberg uncertainty
obey
\be
\label{eq2}
 \hat a_\psi \vert \psi \rangle = 0 \quad .
\ee 
In practice one  fixes the state $ \vert \psi \rangle $ , 
solves the 
{\sl differential equation} 
obtained by equating the expression of Eq.(\ref{neweq2})  to zero and then retains
the solution if it satisfies $ \vert \psi \rangle =  \vert \phi \rangle$.
We use here a harmless abuse of language: assigning numerical values to 
$ \langle \hat x \rangle_\psi  \quad $ and $ \quad  \frac{\langle [\hat x, \hat p] 
\rangle_{\psi} }{2 (\Delta_{\psi} p)^2}  \langle \hat p \rangle_\psi  $ does not
completely fix the state $ \vert \psi \rangle $.

\subsection{The non commutative plane}
  
 Let us consider a two dimensional theory in which the non vanishing
commutation relations are the following \cite{kosi}:
 \be
\label{eq3}
[ \hat x_j , \hat x_k] = i \epsilon_{j k} \, \theta  \quad , \quad [ \hat x_j , \hat p_k] = i
\, \hbar \, \delta_{jk} \quad ; \quad    \theta, \hbar \quad > 0 \quad .
\ee 
This theory admits a representation in which the operators and the scalar
product are given by the following formula
\be
\label{eq4}
\hat x_1 = i \hbar \partial_{p_1} - \frac{1}{2} \frac{\theta}{\hbar} p_2
\quad , \quad \hat x_2 = i \hbar \partial_{p_2} + \frac{1}{2}
\frac{\theta}{\hbar} p_1 \quad  \quad , \quad \hat p_1 = p_1 \quad , \quad
\hat p_2 = p_2  \quad , \quad \quad \langle \phi \vert \psi \rangle = \int
d^2 p \, \phi^*(p) \, \psi(p) \quad . 
\ee   
As  reminded above, the equality in
the Heisenberg relation is attained only for those states which satisfy
Eq.(\ref{eq2}). The operator defined in Eq.(\ref{neweq2}) 
is built from
$\hat x$ and $ \hat p$. Similarly, having two arbitrary operators $\hat u, \hat v$
and a state $\vert \psi \rangle $ one can define a third operator
\be
\hat a = \hat u + i \lambda \hat v + \mu I  \quad {\rm where} \quad 
 \lambda = 
 \frac{\langle [\hat u, \hat v] \rangle_\psi }{2 (\Delta_\psi v)^2}
 \quad {\rm and} \quad \mu = - \langle \hat u \rangle_\psi
 - i \frac{\langle [\hat u, \hat v] \rangle_\psi }{2 (\Delta_\psi v)^2} 
 \langle \hat v \rangle_\psi \quad .
\ee
 In our case, 
\be \label{eq5}
\Delta x_1 \Delta x_2 = \frac{\theta}{2}  \Longrightarrow  {\hat a}_{1\psi}  \vert
\psi \rangle  = 0 \quad , \quad  
\Delta x_1 \Delta p_1 = \frac{\hbar}{2}  \Longrightarrow  {\hat a}_{2\psi}  \vert
\psi \rangle  = 0 \quad , \quad  
\Delta x_2 \Delta p_2 = \frac{\hbar}{2}  \Longrightarrow  {\hat a}_{3\psi}  \vert
\psi \rangle  = 0 \quad ;  
\ee
the operators $\hat a_i$ are given by the following expressions
\be
\label{eq6}
{\hat a}_{1\psi} =  \hat x_1 + i \lambda_1 \hat x_2 + \mu_1 I \quad , \quad 
{\hat a}_{2\psi} =  \hat x_1 + i \lambda_2 \hat p_1 + \mu_2 I \quad , \quad 
{\hat a}_{3\psi} =  \hat x_2 + i \lambda_3 \hat p_2 + \mu_3 I \quad .
\ee
Looking for states $\vert \phi \rangle$ verifying 
${\hat a}_{1\psi} \vert \phi \rangle =  ... = 0 $, the quantities 
$\lambda_i$ and $\mu_i$ become  constants. The second
  and the third
equalities in Eq.(\ref{eq5}) can not be satisfied simultaneously. This is due to the fact 
that if a state satisfying this exists, it should be in the kernel of the two
corresponding operators. Then one would have
 \be
\label{eq7}
[ \hat{a_{2}}_\psi , \hat{a_{3}}_\psi ] \vert \phi \rangle = i
 \theta \vert \phi \rangle = 0 
\ee 
which  admits only the null vector as a solution. A more extensive analysis
of the other equalities can be found in \cite{kosi}.

The "constants" $\mu_k$ appearing in the operators $\hat a_i$ in Eq.(\ref{eq6}) do not
play any role in the considerations implying the commutators. We shall discard
them for simplicity in the   model  considered below.
For simplicity, the real quantities $\lambda_k$ will be treated as "constants" in what
follows; that means the state $\vert \psi \rangle$ has been fixed and one is 
solving the equation for $\vert \phi \rangle$. Due to this convention, we shall
not write a subscript to our operators explicitly. 

\subsection{The two dimensional model with a minimal uncertainty in length.}
   The first extension of the one dimensional model endowed with a minimal
length uncertainty was proposed in \cite{A_2}:  \be
\label{eq8}
[ \hat x_j ,\hat p_k] = i \hbar (1 + \beta  \hat{{\vec p}^2} ) \delta_{j k} 
\quad , \quad [ \hat x_j ,\hat x_k] = 2 i \hbar \beta ( \hat p_j  \hat x_k -
\hat p_k  \hat x_j)  \quad . 
\ee 
It lacks translational  symmetry since the second relation is not invariant
under the transformation \mbox{$\hat x_k  \rightarrow \hat x_k + \hat \alpha_k I,
\,\, \hat p_k  \rightarrow \hat p_k $}.
The second extension of the KMM model to higher dimensions is invariant under
rotations and translations. The non trivial commutation relations are  \cite{kmm3}:

\be
\label{eq12}
[ \hat x_j , \hat p_k] = i \hbar \left(  f(\hat{{\vec p}^2})  \delta_{j
k} + g(\hat{{\vec p}^2}) \hat p_j  \hat p_k \right) \quad , \ee 
and the condition 
\be
\label{eq13}
g = \frac{2 f f'}{f - 2 p^2 f'} \quad .
\ee
enforces the commutation of the positions among themselves. The same property holds for 
the momenta.  
The functions $f$ and $g$ are supposed positive. In this model, a position operator
 in one
direction has a non trivial commutation relation with  the momentum associated
to another direction:
\be  
\label{eq14} 
[ \hat x_1 , \hat p_2 ] = i h g(\hat{{\vec p}^2}) \hat p_1  \hat p_2 \quad .
\ee 
A representation of this extension is realized by the following formulas:
\be
\label{eq15}
\hat x_i = i \hbar \left( f' + {\vec p}^2  g'+ \frac{3}{2} g \right) p_i + f
\partial_{p_i} + g p_i p_j  \partial_{p_j} \quad , \quad \hat p_i = p_i \quad
, \quad \langle \phi \vert \psi \rangle = \int d^2 p \phi^*(p) \psi(p) \quad .
\ee 

In a space of dimension two, this theory has four non trivial uncertainty relations concerning the
following couples of variables: $(x_1,p_1),(x_2,p_2),(x_1,p_2),(x_2,p_1)$. The
 corresponding operators are
\be
\label{eq16}
\hat a_1 = \hat x_1 + i \lambda \hat p_1 \quad , \quad 
\hat a_2 = \hat x_2 + i \mu \hat p_2 \quad , \quad 
\hat a_3 = \hat x_1 + i \tau \hat p_2 \quad , \quad 
\hat a_4 = \hat x_2 + i \sigma  \hat p_1 \quad , 
\ee
so that the relevant commutators, are, in this case

\begin{eqnarray}
\label{eq17}
&& [ \hat a_1 , \hat a_2 ] = \hbar g(\hat{{\vec p}^2}) ( - \mu \hat p_1 \hat
p_2 + \lambda \hat p_2 \hat p_1) \quad , \quad 
[ \hat a_1 , \hat a_3 ]= - \hbar \left( g(\hat{{\vec p}^2}) ( \tau \hat p_1
\hat p_2 - \lambda {\hat p_1}^2 ) - \lambda  f(\hat{{\vec p}^2})  \right) 
\quad , \quad  \nonumber\\
&&[ \hat a_1 , \hat a_4 ]= - \hbar \left( g(\hat{{\vec p}^2}) ( - \lambda \hat
p_2 \hat p_1 + \sigma {\hat p_1}^2 ) + \sigma  f(\hat{{\vec p}^2})  \right) 
\quad , \quad 
[ \hat a_2 , \hat a_3 ]= - \hbar \left( g(\hat{{\vec p}^2}) ( - \mu \hat
p_1 \hat p_2 + \tau {\hat p_2}^2 ) + \tau  f(\hat{{\vec p}^2})  \right) 
\quad , \quad  \nonumber\\
&&[ \hat a_2 , \hat a_4 ]= - \hbar \left( g(\hat{{\vec p}^2}) (  \sigma \hat
p_2 \hat p_1 - \mu {\hat p_2}^2 ) - \mu  f(\hat{{\vec p}^2})  \right) 
\quad , \quad 
[ \hat a_3 , \hat a_4 ]= - \hbar \left( g(\hat{{\vec p}^2}) (  \sigma {\hat
p_1}^2 -  \tau {\hat
p_2}^2 ) +  (\sigma - \tau) f(\hat{{\vec p}^2}) \right)  \quad .
\nonumber\\ 
\end{eqnarray} 

From now on, we use the  form of the operators given earlier.
The first commutator has a non trivial kernel for $\lambda=
\mu$. In particular, there are non zero vectors $\vert \psi \rangle$ fulfilling
$\lambda=\mu$ which also fulfill $\hat a_{2_\psi} \vert \psi \rangle =
\hat a_{3 \psi} \vert \psi \rangle = 0$, and  saturate simultaneously the uncertainty
relations concerning the couples of variables $(x_1,p_1)$ and
$(x_2,p_2)$. These states are the ones used by A.Kempf \cite{A_2}, S.Detournay 
Cl.Gabriel and Ph.Spindel 
 \cite{spindel} while studying the existence of a minimal uncertainty
in length. We are interested in a
conceptually different problem: {\em in this work we put all the non trivial
commutation relations on the same footing.}

The
action of the second commutator on a state $\vert \phi \rangle$ vanishes 
only if the equation
\be
\label{eq18}
\left( ( \tau p_1 p_2 - \lambda p_1^2 ) - \lambda \frac{f({\vec p}^2)}{f({\vec
p}^2)} \right) \phi(\vec p) = 0  \ee 
is satisfied. Clearly this is the case only when the expression under
parentheses vanishes. As its second term is a function of $ {\vec p}^2$,
 one should have the same thing for its first term. This is possible
only if $\tau = \lambda=0 $, but  since at the end of the analysis
\be  
\label{eq19} 
\lambda = \frac{\hbar}{2(\Delta p_1)^2} \int  d^2 p \phi^*(\vec p)
\left(f({\vec p}^2) + g({\vec
p}^2) p_1^2 \right) \phi(p)  \quad ,
\ee 
and the functions $f,g$ are positive, this is excluded.

\section{Periodicity of the harmonic oscillator}

The coherent states, in the usual case, are not only the states which  
saturate simultaneously all the non trivial uncertainties; they also reproduce the classical behavior
of the harmonic oscillator. In this section we will interest ourselves to  the mean values of the positions and momenta of
a harmonic oscillator in the theories under study. In  usual quantum theory, any 
state which is a solution of the Schrodinger equation for the harmonic oscillator
displays periodic positions and momenta; moreover the period coincides exactly with the
one found in classical mechanics \cite{cohen}. We will see that this is generically not the case
when non commutativity sets in.

\subsection{The non commutative plane}

Working in the
 Heisenberg picture, the states are time independent while the operators vary according
 to the evolution  equation 
\be
\label{eq94}
\left(
 \begin{array}{c}
  \dot{\hat x}_1   \\ \dot{\hat x}_2 \\ \dot{\hat p}_1 \\  \dot{\hat p}_2  
\end{array}  
\right) = 
\left(
 \begin{array}{cccc}
  0                         & \frac{k \theta}{\hbar} & \frac{1}{m} & 0 \\
  - \frac{k \theta}{\hbar}  &            0           &          0  & \frac{1}{m} \\
  - k                       &            0           &          0  &    0        \\
  0                         &           - k          &          0  &    0       
 \end{array}
\right)
\left(
 \begin{array}{c}
 {\hat x}_1  \\ {\hat x}_2 \\ {\hat p}_1 \\  {\hat p}_2   
\end{array} \right) \quad .
\ee 
This system of differential equations is easily solved using the exponential method.
For example, one has for the first position
\be
\label{eq95}
 \hat x_1(t)   = M_{11}(t)   \hat{x}_1(0)  +
  M_{12}(t)  \hat{x}_2(0)  + 
 M_{13}(t)  \hat{p}_1(0)  +  M_{14}(t)  \hat{p}_2(0)  \quad ,
\ee 
with
\begin{eqnarray} 
\label{eq96} 
M_{11}(t) & = & \frac{1}{\lambda_1 + \lambda_2} 
( \lambda_1 \cos{\lambda_1 t} +  \lambda_2 \cos{\lambda_2 t} ) \quad , \quad
M_{12}(t)  =  \frac{1}{\lambda_1 + \lambda_2} 
(- \lambda_1 \sin{\lambda_1 t} +  \lambda_2 \sin{\lambda_2 t} ) \nonumber \\
M_{13}(t) & = & \frac{\lambda_1 \lambda_2}{\lambda_1^2 - \lambda_2^2} \frac{\theta}{\hbar}
( - \sin{\lambda_1 t} - \sin{\lambda_2 t} ) \quad , \quad
M_{14}(t) = \frac{\lambda_1 \lambda_2}{\lambda_1^2 - \lambda_2^2} \frac{\theta}{\hbar}
( - \cos{\lambda_1 t} + \cos{\lambda_2 t} ) \quad .
\end{eqnarray} 
The eigenvalues of the matrix  appearing
in Eq.(\ref{eq94}) are  $\pm\, i \lambda_1 {\rm and } \, \pm i \lambda_2$ where
\begin{equation}
\label{eq97}
\lambda_{1} =  \sqrt{ \frac{k}{m} + \frac{k^2 \theta^2}{2 \hbar^2} - 
            \frac{k^{3/2} \theta}{2 \hbar^2 \sqrt{m}}  
	    \sqrt{4 \hbar^2 + k m \theta^2}} \quad , \quad
\lambda_{2} =  \sqrt{ \frac{k}{m} + \frac{k^2 \theta^2}{2 \hbar^2} + 
            \frac{k^{3/2} \theta}{2 \hbar^2 \sqrt{m}}  
	    \sqrt{4 \hbar^2 + k m \theta^2}}	    
\end{equation} 
are real numbers. For a given
wave function, the operators evaluated at the initial time 
(like $ \langle \hat x_1(0) \rangle $) are obtained thanks to the 
expressions displayed in Eq.(\ref{eq4}). The mean value of the first position in any 
state $\psi(p_1,p_2)$ can be rewritten as
\be
\label{eq98}
\langle  \hat{x}_1(t) \rangle  =  \frac{1}{\hbar (\lambda_1 - \lambda_2) 
(\lambda_1 + \lambda_2)} ( c_1 \cos{\lambda_1 t} + s_1 \sin{\lambda_1 t} +
 c_2 \cos{\lambda_2 t} + s_2 \sin{\lambda_2 t} ) \quad ,
\ee
with
\begin{eqnarray}
\label{eq99}
 c_1 &=&   \hbar \langle \hat x_1(0) \rangle \lambda_1^2 - \lambda_1 \lambda_2
   ( \hbar \langle \hat x_1(0) \rangle - \theta \langle \hat p_2(0) \rangle ) 
   \quad , \quad
s_1 =   - \hbar \langle \hat x_2(0) \rangle \lambda_1^2 + \lambda_1 \lambda_2
   ( \hbar \langle \hat x_2(0) \rangle - \theta \langle \hat p_1(0) \rangle ) \quad ,
    \nonumber\\    
c_2 &= &  \hbar \langle \hat x_1(0) \rangle \lambda_2^2 - \lambda_1 \lambda_2
   ( \hbar \langle \hat x_1(0) \rangle + \theta \langle \hat p_2(0) \rangle ) 
  \quad , \quad   
s_2 =  \hbar \langle \hat x_2(0) \rangle \lambda_2^2 + \lambda_1 \lambda_2
   (- \hbar \langle \hat x_2(0) \rangle + \theta \langle \hat p_1(0) \rangle ) 
   \quad .
\end{eqnarray}  

From Eq.(\ref{eq98}), one sees that the position mean value will  be 
periodic for all states only 
if the ratio of the two frequencies $\lambda_1/\lambda_2$ is rational. This requires 
a fine tuning of the parameters and so is not generic. 
In the cases where the ratio 
is not rational, there are states  for which the coefficients  are such that the 
components of one of the frequencies vanish. For example, the component of frequency
 $\lambda_1$ 
disappears in $ \langle \hat x_1(t) \rangle$  if the following conditions are satisfied:
\be
\label{eq100}
\theta \lambda_2  \langle \hat p_{1}(0) \rangle + \hbar (\lambda_1 - \lambda_2 ) 
 \lambda_2  \langle \hat x_{2}(0) \rangle = 0 \quad , \quad 
 \theta \lambda_2 \langle \hat p_{2}(0) \rangle + \hbar (- \lambda_1 + \lambda_2 ) 
 \lambda_2 \langle \hat x_{1}(0) \rangle = 0 \quad .
\ee 
Moreover, these two conditions also suppress the $\lambda_1$ components in the mean values
of the remaining variables
$x_2,p_1,p_2$. The above conditions translate into the vanishing of the following
integrals:
\be
\label{eq101}
\int d^2 p \psi^\ast(p) \left( i \hbar^2 (\lambda_1 - \lambda_2) \partial_{p_2} + 
\frac{1}{2} \theta (\lambda_1 + \lambda_2 ) p_1  \right) \psi(p) \quad , \quad
\int d^2 p \psi^\ast(p) \left( i \hbar^2 (\lambda_1 - \lambda_2) \partial_{p_1} - 
\frac{1}{2} \theta (\lambda_1 + \lambda_2 ) p_2  \right) \psi(p) \quad .
\ee
It is readily found that the operators contained in the parentheses do not have a 
common zero eigenvalue so that the states we are looking for cannot be found by solving two 
first order differential equations. However, one can show that the wave function
\be
\label{eq102}
\psi(p) = \exp{( (- a_1^2+i a_2) p_1^2 + ( - b_1^2+i b_2) p_2^2+ (c_1+i c_2) p_1 +
(d_1 + i d_2) p_2)}
\ee
satisfies the two conditions provided that the following relations hold:
\be
\label{eq103}
c_2 = - \frac{a_2 c_1}{a_1^2} - \frac{1}{4}  \frac{\theta}{ \hbar^2} 
\frac{\lambda_1+\lambda_2}{\lambda_1-\lambda_2} \frac{d_1}{b_1^2} \quad , \quad
d_2 = - \frac{b_2 d_1}{b_1^2} + \frac{1}{4}  \frac{\theta}{ \hbar^2} 
\frac{\lambda_1+\lambda_2}{\lambda_1-\lambda_2} \frac{c_1}{a_1^2} \quad .
\ee
It can be easily seen from Eqs.(\ref{eq97},\ref{eq98}) that the usual theory is recovered when the 
deformation parameter goes to zero.

\subsection{The model possessing a minimal length uncertainty}

For the KMM theory, the situation is more complicated and we will treat only the 
one dimensional case. The analysis in the Heisenberg picture is not straightforward 
because the evolution of the operators leads to  non linear equations like
\be
\label{eq104} 
\dot{\hat{x}} = \frac{1}{m} ( 1+ \beta \hat{p}^2 ) \hat p \quad . 
\ee
We will rather use the Schrodinger picture and the knowledge of the spectrum of the 
harmonic oscillator \cite{A_2}. The operators are now fixed and the time dependence is
carried by the state. The energy eigenstates form a basis of the Fock space and so an 
arbitrary state can be expanded as $ \vert \psi \rangle = \sigma^n \vert n \rangle. $
The mean value of the position in this state reads
\be
\label{eq105}
\langle \hat x(t) \rangle = \sum_{m,n} \sigma^n (\sigma^*)^m 
\langle m \vert \hat x \vert n \rangle  \exp{\left(\frac{i}{\hbar} 
( E_m - E_n ) t \right) } \quad ;
\ee 
the time dependence is entirely contained in the exponentials. In the usual case the matrix element
 $ \langle m \vert \hat x \vert n \rangle$ is non vanishing only for integers $m,n$
  which differ by one unit. As the energy spectrum is then linear, the arguments of 
  the  exponentials are all equal if we discard the signs. Collecting them, one ends up
  with the usual superposition of sine and cosine functions. In the KMM theory, the spectrum of the 
  harmonic oscillator is quadratic \cite{A_2}: $E_n = a n^2 + b n + c $, where
  the constants $a,b,c$ depend on the characteristics of the oscillator
  and the deformation parameter. This leads us to 
  the following expression for the position mean value:
  \be
  \label{eq106}
  \langle \hat x(t) \rangle = \sum_{m,n} \tau_{m,n} ( \cos{\omega_{m n}t} + 
  i \sin{\omega_{m n}t} ) \quad , \quad \omega_{m n} = 
  \frac{1}{\hbar} ( m - n) ( a ( m+n) + b ) \quad .
  \ee
This mean value will be periodic only if  the frequency ratio 
$\omega_{m_1 n_1}/\omega_{m_2 n_2}$ is a rational number every time
$ \tau_{m_1,n_1},\tau_{m_2,n_2}$ are non vanishing. Then, there must exist a rational number $r_{12}$
such that
\be
\label{eq107}
r_{12} = \frac{a(m_1+n_1) +b}{a(m_2+n_2) +b} \quad .
\ee
From this one infers that $b/a$ must then be rational; replacing the quantities $a$ 
and $b$ in terms of the quantum parameters \cite{A_2}, this amounts to ask that
\be
\label{eq108}
\frac{\beta m \hbar \omega}{ \beta m \hbar \omega + \sqrt{4 + \beta m \hbar \omega}}
\ee
is rational. When this condition is not satisfied, periodicity is lost.

\section{The least square variational principle in the usual 1D case}
  
We have seen in the last sections that two defining properties of the coherent
states(reproduction of the harmonic oscillator's classical behavior, saturation of the
non trivial Heisenberg uncertainties) are lost as one goes to non commutative
theories. The question we now face  is: what possible generalization can one 
adopt? The first possibility relies on a deformation of the algebra generated by 
the creation-destruction operators: we will say a few words about it in the 
next section. The generalization we propose here strictly relies on the 
uncertainties. Let us consider a theory whose position and momentum operators are generically
 denoted $\hat G_i$ and whose non trivial uncertainty relations can not be satisfied
simultaneously. 

{\sl We will call quasi classical or squeezed states those which display
a minimum of a functional of the type}
\be 
\label{eq22} 
S_{n,m}=\sum_{j,k} \left| (\Delta  G_j)^{2 n} (\Delta  G_k)^{2 n} -  \frac{ \vert
\langle [\hat G_j, \hat G_k ] \rangle  \vert^{2 n}}{2^{2 n}} \right|^{2 m}  \quad .
\ee 
The sum runs on all couples  $(j,k)$  such that the  commutators
$[ \hat G_j, \hat G_k ] $ do not
 vanish. The values of the indices $(n,m)$ 
 in which we
will  be interested are $1/2$ and $1$; we shall come back to this point
later. It is important  that all  choices of $(n,m)$ lead essentially to the same
differential equation when the action is varied, as we shall see. We restrict
ourselves to pairs of variables with non vanishing commutators
in order to recover the desired results in the unmodified theory for higher
dimensions; this will be discussed at the end of this section.

In the usual theory, the action simply
vanishes on  squeezed states. 
The difficult point in our proposal lies in the fact that the variation of the
sum given in Eq.(\ref{eq22}) leads to a second order differential equation
which in principle admits more solutions. Let us show how this works in the
simplest i.e undeformed theory. The action we consider is
\be
\label{eq23}
S = \left| (\Delta x)^2  (\Delta p)^2 - \frac{1}{4}  \right|
  \quad ,
\ee 
where for simplicity we have set $\hbar=1$.
{\sl The Heisenberg inequality tells us that we can remove the absolute value}.
The definition of the uncertainty
\be 
\label{eq24} 
(\Delta x)^2 = \frac{\langle \psi \vert \hat x^2  \vert  \psi\rangle}{\langle
\psi \vert   \psi\rangle} - \left( \frac{\langle \psi \vert \hat x  \vert 
\psi\rangle}{\langle \psi \vert   \psi\rangle} \right)^2
\ee 
renders necessary the computation of  derivatives such as
\be
\label{eq25}
\frac{\delta}{\delta \psi^*(p)} \langle \psi \vert  \hat x  \vert \psi\rangle =
\frac{\delta}{\delta \psi^*(p)}  \int dq \, \psi^*(q) i \partial_q \psi(q) =
\int dq \, \delta(p-q) i \partial_q \psi(q) = i \partial_p \psi(p)   \quad . \ee 
Performing similar computations, we obtain the following equation when varying the action $S$:
\be
\label{eq26}
{\cal O} \vert \psi \rangle = \langle \psi \vert \psi \rangle \frac{\delta S}{\delta \psi^\ast(p)}
 =  a   \psi^{''}(p) + i b \psi^{'}(p) + ( c p^2 + d p + e) \psi(p) = 0  \quad ,
\ee 
where $a,b,c,d,e$ are linked to the observables of the
solution by the relations
\begin{eqnarray}
\label{eq27}
a &=& - (\Delta p)^2  
\quad , 
\quad b = -2  (\Delta p)^2 \langle \hat x \rangle
\quad , 
\quad c = (\Delta x)^2  \quad , \quad
d = -2  (\Delta x)^2 \langle
\hat p \rangle  \quad , \nonumber\\
 e &=& - 2 (\Delta x)^2  (\Delta p)^2 + (\Delta x)^2  \langle \hat p \rangle^2
 + (\Delta p)^2  \langle \hat x \rangle^2 \quad .
\end{eqnarray}
The mean values are taken on the same state $\vert \psi \rangle$.
The appearance of the observables of the solution such as $\langle \hat x \rangle,
\cdots $ in this equation is similar to the situation encountered in
Eq.(\ref{eq2}) i.e for  usual coherent  states. One
can similarly to  Eq.(\ref{neweq2}) define an operator $ {\cal O_\psi} $
 depending
on a state $\vert \psi \rangle$ such that its action on any state 
$\vert \phi \rangle$ is given  by 
\begin{eqnarray}
\label{eq1000}
{\cal O_\psi} \vert \phi \rangle &=&  - (\Delta_\psi p)^2  \,\,  \phi^{''}(p)
 -  2 \, i \,
  (\Delta_\psi p)^2 \, \langle \hat x \rangle_\psi \, \, \phi^{'}(p) \nonumber\\
  & + & 
  \left[  (\Delta_\psi x)^2 p^2 -2  (\Delta_\psi x)^2 \langle
\hat p \rangle_\psi \,  p +   (- 2 (\Delta_\psi x)^2  (\Delta_\psi p)^2 + 
(\Delta_\psi x)^2  \langle \hat p \rangle_\psi^2
 + (\Delta_\psi p)^2  \langle \hat x \rangle_\psi^2 ) \right] \,\, \phi(p)
 \quad ,
\end{eqnarray}
leading to a differential equation.
 
 Let us  verify that the well known coherent  states 
satisfy  our equation; we  look for  Gaussians \cite{cohen}:
\be
\label{eq28}
\phi(p) = N \exp{( (z_1+ i z_2) p^2 + (z_3+i z_4) p)} 
\ee 
which are solutions of
\be
\label{eq1001}  
 {\cal O_\psi} \vert \phi \rangle = 0 \quad .
\ee
 This happens if the  $z_i$ are given by
 \be 
\label{eq29} 
z_1= - \sqrt{- \frac{ c}{4 a} } \equiv - 
\sqrt{- \frac{ (\Delta_\psi x)^2}{4 (\Delta_\psi p)^2} } \quad , \quad z_2 = 0 \quad , 
\quad z_3 = \frac{d}{2 a} \sqrt{- \frac{a}{c} } \quad , \quad
z_4 = - \frac{b}{2 a} 
\ee 
and the following relation between the parameters $a,...,e$ holds:
\be
\label{eq1004}
e = - \frac{b^2}{4 a} - \sqrt{-a c} + \frac{d^2}{4 c} \quad .
\ee
The final expression of $z_1$ in Eq.(\ref{eq29}) is obtained using 
Eq.(\ref{eq27}); the same thing can be done for the other $z_k$.
Due to Eq.(\ref{eq27}), the $z_i$ are real. They  depend on the state 
$\vert \psi \rangle $. The normalization of the wave function to
one  is achieved by an appropriate choice of the pre factor $N$. One finds
 \be
\label{eq31}
\Delta_\phi x =  \frac{1}{\sqrt{2}} \left(- \frac{ c}{a} \right)^{1/4} \quad , 
\quad \Delta_\phi p = \frac{1}{\sqrt{2}} \left(- \frac{ a}{c} \right)^{-1/4}
  \quad .
\ee 
Imposing that the states $ \vert \phi \rangle$ and $  \vert \psi \rangle$
coincide amounts to replace $\Delta_\phi x$ by $ \Delta_\psi x$ in the left
sides of Eq.(\ref{eq31}) and the parameters $a,...,e$ by their expressions 
given by Eq.(\ref{eq27}). This leads to the equality 
\be 
\label{eq1002}
 \Delta_\psi x \Delta_\psi p = \frac{1}{2} \quad .
\ee
This relation could also be obtained by multiplying the two equalities given in
Eq.(\ref{eq31}) or by developing  Eq.(\ref{eq1004}) using Eq.(\ref{eq27}).
The action $S$ vanishes on the state we have obtained. We have thus explicitly
verified that the usual coherent states are captured by our procedure.

Instead of simply looking for the conditions under which the known Gaussians
obey  Eq.(\ref{eq26}), one may look for the
cases in which  the operator of Eq.(\ref{eq1000}) factorizes, leading to two first order operators:
 ${\cal O_\psi}= \hat k_1 \hat k_2$. It is not difficult to show that this occurs 
when Eq.(\ref{eq1004}) is satisfied. One then has
\begin{eqnarray}
\label{eq1005}
\hat k_1 \phi(p) &=& - i \sqrt{-a} \phi'(p) + \left(
\frac{1}{2} \left( - \frac{b}{\sqrt{-a}} + \frac{i d}{\sqrt{c}} \right) 
+ i \sqrt{c} p \right) \phi(p) \nonumber\\
\hat k_2 \phi(p) &=& - i \sqrt{-a} \phi'(p) + \left(
\frac{1}{2} \left( - \frac{b}{\sqrt{-a}} - \frac{i d}{\sqrt{c}} \right) 
- i \sqrt{c} p \right) \phi(p) \quad .
\end{eqnarray}
A state which is in the kernel of $\hat k_2$ is obviously in the kernel of
${\cal O}$. Using the expressions of the operators $\hat x, \hat p$, one finds
\be
\label{eq1006}
\hat{k_2} = (\Delta_\psi p) \left[ \hat x - \langle \hat x \rangle_\psi I + 
i \left( \frac{\Delta_\psi x}{\Delta_\psi p} \right)  (\hat p - 
\langle \hat p \rangle_\psi I ) 
\right] \quad .
\ee
The relation given in Eq.(\ref{eq1004}) which allows  factorization also ensures
Eq.(\ref{eq1002}) as stated before. This can then be used to show that the 
operator $ \hat{k_2}$ is proportional to the first order operator
 appearing in Eq.(\ref{neweq2}). We have thus seen how the usual first order
differential equation which leads to squeezed states
in the undeformed case can be recovered with the method we propose.
 
So far we have studied very specific solutions to the second order equation we
obtained. To find the
most general solution to Eq.(\ref{eq26}), let us make the following 
change of variables:
\be 
\label{eq34} 
 \psi(p)= \exp{\left( \frac{- 4 i b p+ (-a)^{1/2} c^{-3/2} (d+2 c p)^2 }{8 a} \right)}
  (d+2 c p) X(q)  \quad ; \quad q= - \frac{1}{4}(- a)^{-1/2} c^{-3/2} (d+2 c p)^2 \quad.
\ee
To simplify the expressions, let us also introduce the quantity 
\be
\label{eq35}
\alpha= \frac{ -(-a)^{1/2} b^2 c +
 a (12 a c^{3/2}+ (-a)^{1/2}(d^2 - 4 c e ) )}{16 a^2 c^{3/2}} \quad .
\ee 
The differential equation now becomes
\be
\label{eq36}
X''(q) + \left(1 + \frac{3}{2 q} \right) X'(q) + \frac{\alpha}{q} X(q) = 0 
\ee
and its general solution can be   written as a sum of hyper geometric functions:
\be
\label{eq37}
X(q) = C_1 \,  q^{-1/2} \, \,   {_1} F_{1} \left( -\frac{1}{2}+ \alpha, \frac{1}{2},- q \right)+
 C_2 \, \, \,  {_1} F_1 \left( \alpha, \frac{3}{2},- q \right) \quad .
\ee
Using for the constant $e$ the expression displayed in Eq.(\ref{eq1004}), the
first argument of the two hyper geometrics equals $1/2$. As the second
hyper geometric becomes a constant in this case, one recovers the Gaussian
solution given in Eq.(\ref{eq28}) by taking $C_2=0$.

Apart from the states which are known to have a vanishing
value of $S$, the formalism we have used introduces new
states. The fact that the Gaussians displayed
in Eq.(\ref{eq28}) exhibit the absolute minimum of the action and form an over
complete set will be enough to discard the other states.

It should be stressed that normalizable solutions do not exist for all the values
of the parameters. In fact, the change of variables
\be
\label{eq38}
\psi(p) = \exp{\left( - \frac{i b}{2 a} p \right)} u(p)
\ee
gives to Eq.(\ref{eq26}) the following form
\be
\label{eq39}
  u''(p) + V(p) u(p) = 0   \quad , \quad 
 V(p) =  \left( \frac{c}{a} p^2 + \frac{d}{a} p + \frac{e}{a} + \frac{b^2}{4 a^2} 
  \right) \quad .
\ee
Integrating by parts and discarding the boundary term, one obtains the relation
\be
\label{eq40}
\int_{-\infty}^{+\infty} dp  \left( u'(p)^2 - V(p) u(p)^2 \right) = 0 \quad ,
\ee
which  can not be satisfied unless the opposite of $ V(p)$ admits two distinct zeros.
This results 
in the following inequality:
\be
\label{eq41}
e >\frac{d^2}{4 c} - \frac{b^2}{4 a} \quad .
\ee
This  reasoning applies when the function $u(p)$ is real. If it is complex,
its real and imaginary parts obey  Eq.(\ref{eq39}) whose parameters are real;
the same conclusion holds.

To end this section, let us first notice that the choice $n=m=1/2$ in
Eq.(\ref{eq23}) can be chosen without changing drastically the situation. In
fact, trivial relations such as
\be  
\label{eq42}  
  \frac{\delta}{\delta \psi^{*}(p)} \Delta x = \frac{1}{2 \Delta x  }
\frac{\delta}{\delta \psi^{*}(p)} (\Delta x)^2  
\ee 
show that if we work with $n=1/2$, we will end up with  a
differential equation of the same form than  Eq.(\ref{eq26}). The "only" difference will
be encoded in a modified Eq.(\ref{eq27}) which gives the
link between the parameters $a,\cdots,e$ and the observables of the state.
What happens when one considers a different value of $m$? Then, the quantities $a,...,e$
of Eq.(\ref{eq27}) are multiplied by the same factor 
$2 m ((\Delta_\psi x)^{2 n} (\Delta_\psi p)^{2 n} - 1/4^n )^{2 m -1}$ {\sl which vanishes} for
the usual coherent states. The differential equation (\ref{eq1000})  then
reduces, for this 
choice of $\vert \psi \rangle$, to the identity $0 = 0 \, \phi(p)$.
Nevertheless, the extra factor being common to all of the parameters $a,...,e$,
one can divide by it and obtain an equation which makes sense and leads to
the  states obtained above. For most of this work we shall take $m=1/2$ for 
simplicity. 

Let us now consider how our proposal applies to higher dimensions in the 
unmodified case. Taking the action to be 
\be
\label{eq23bis}
S = \sum_{j=1}^N \left| (\Delta x_j)^2  (\Delta p_j)^2 - \frac{1}{4}  \right|
  \quad ,
\ee 
one finds  the equation
\be
\label{nontri}
\sum_{j=1}^N {\cal O}_{x_j,p_j} \vert \phi \rangle = 0 \quad ,
\ee              
where the operators ${\cal O}_{x_j,p_j}$ are given by Eq.(\ref{eq1000}). 
As these $N$  operators commute,  particular solutions are found
in the intersection of their kernels. These are the usual coherent states 
in $N$ dimensions. One can see from here that allowing in the action terms 
associated to couples like $\hat x_1, \hat p_2$ which commute would spoil this result.

\section{High dimensional extensions}

\subsection{Generalized coherent states}
    
The generalization of the notion of coherent states developed by \cite{pere} 
for non commutative
    theories has been considered by many authors \cite{coh1,coh2,coh3,coh4}.
 Having a
     deformation of the
     usual position-momentum commutation relations, one constructs operators
     obeying the relation
\be
\label{eq43}
[  \tilde a, \tilde a^{+} ] = F(\tilde a \tilde a^{+}) \quad .
\ee
The coherent states are then defined to be the eigenstates of the modified 
destruction operators:
\be
\label{eq44}
 \tilde a \vert \xi \rangle = \xi \vert \xi \rangle \quad .
 \ee  
The function $F$ is constant in the usual case. When the deformed
 creation-destruction
operators can be constructed from the usual ones ($a, a^{+}$) in the following way:
\be
\label{eq45}
\tilde a = f( \hat n +1) a  \quad , \quad n = a^{+} a \quad ,
\ee
 a link exists between the functions $f$ and $F$.
This approach is mathematically   useful in the sense that a 
decomposition of the unity is obtained quite easily and  many properties of the usual 
coherent states survive. However, the link with the Heisenberg uncertainties 
is far from trivial. 

One can  define a deformed creation-destruction algebra, find the eigenstates
of the destruction operators, compute the associated uncertainties and then try
to see to which extent they are minimal. We here address the problem in a different way: we impose the minimization of
the uncertainties and try to find the relevant states. This approach is more
analytical than algebraic. In the usual  case, the two
constructions  lead to the same result as shown above.

To avoid any confusion, the states obtained by our procedure will be called
{\sl quasi classical} or {\sl squeezed} while the ones obtained trough a deformation of
 the destruction operator will be referred to as {\sl coherent}.

\subsection{The Non commutative plane}

 Let us define  the  variables $\hat X_k, \hat P_k$ 
 linked to the ones given in Eq.(\ref{eq3}) by the relations  
\be
\label{eq46}
\hat x_k = \sqrt{\theta} \hat X_k  \quad , \quad  \hat p_k =
\frac{\hbar}{\sqrt{\theta}} \hat P_k  \quad .
 \ee 
Illustrating now our method with the values $m=n=1/2$, the three non trivial
commutation relations lead to the dimensionless  action
\be 
\label{eq47} 
\tilde S = \left| \Delta X_1 \Delta X_2 - \frac{1}{2}  \right| + \left|
\Delta X_1 \Delta P_1 - \frac{1}{2}  \right| +  \left| \Delta X_2 \Delta P_2
- \frac{1}{2} \right|   \quad . 
\ee
Multiplying by an overall factor, and going back to the former variables, we
obtain the action which will be used in this subsection:
 \be
\label{eq48}
S=  \hbar \left| \Delta x_1 \Delta x_2 - \frac{1}{2} \theta
\right| + \theta \left| \Delta x_1 \Delta p_1 - \frac{1}{2}  \hbar
\right| +  \theta \left| \Delta x_2 \Delta p_2 - \frac{1}{2} 
\hbar \right|   \quad .  
\ee 
From a dimensional analysis, one could  guess the form of the
terms appearing in Eq.(\ref{eq48}) but not their relative weights. The
introduction of generic dimensionless variables and the adoption of the
democratic rule displayed in Eq.(\ref{eq47}) fixes in an unambiguous way the
action displayed in Eq.(\ref{eq48}). Introducing
constant factors to balance the different terms  leads essentially to the same
differential equation.

The discussion of section $2$ has shown that  there is no state on which 
this action vanishes. We now look for those on which it is extremal. The absolute
values can be removed as in the previous section. Varying
the action we obtain the field equation
\be
\label{eq49}
\langle \psi \vert \psi \rangle \frac{\delta S}{\delta \psi^*(p)} = b_1 \frac{\delta }{\delta \psi^*(p)}
\Delta x_1 + b_2 \frac{\delta }{\delta \psi^*(p)} \Delta x_2 + b_3 \frac{\delta
}{\delta \psi^*(p)} \Delta p_1 + b_4 \frac{\delta }{\delta \psi^*(p)} \Delta
p_2 = 0  \quad ,
\ee 
where the coefficients $b_i$ are given by the following expressions
\be
\label{eq50}
b_1 =  \hbar    \Delta x_2 +  \theta  
  \Delta p_1  \quad , \quad
 b_2 =  \hbar   \Delta x_1 +  \theta
   \Delta p_2  \quad , \quad
b_3 =  \theta   \Delta x_1 \quad , \quad  
b_4 =  \theta   \Delta x_2 \quad .   
\ee
Intermediate results such as
\be
\label{eq51}
\frac{\delta }{\delta \psi^*(p)} \Delta x_1^2 = \left[ - \hbar^2
\partial^2_{p_1} - i ( \theta p_2 + 2 \hbar \langle \hat x_1 \rangle)
\partial_{p_1}  + \left(  \frac{1}{4} \frac{\theta^2}{\hbar^2}  p_2^2 +
\frac{\theta}{\hbar} \langle \hat x_1 \rangle p_2 + \langle \hat x_1 \rangle^2
- (\Delta x_1)^2 \right) \right] \psi(p) \quad ,
\ee 
\be
\label{eq52}
\frac{\delta }{\delta \psi^*(p)} \Delta p_1^2 = \left( p_1^2 - 2 \langle
\hat p_1\rangle  p_1 + \langle
\hat p_1\rangle^2 - (\Delta p_1)^2 \right) \psi(p) \quad ,
\ee 
are necessary to recast  the equation in the following form
\be 
\label{eq53} 
\left[ \bar{a}_1 \partial^2_{p_1} + \bar{a}_2 \partial^2_{p_2} + i 
 \left( \bar{a}_1
\frac{\theta}{\hbar^2}  p_2 + \bar{a}_3 \right)
\partial_{p_1}  + i  \left( -\frac{\theta}{\hbar^2} \bar{a}_2 p_1
+ \bar{a}_4 \right) \partial_{p_2} + ( \bar{a}_5 p_1^2 + \bar{a}_6 p_2^2
+ \bar{a}_7 p_1 + \bar{a}_8 p_2 + \bar{a}_9 ) \right] \psi(p) = 0  \quad . 
\ee  
The real numbers
$\bar{a}_i$ are linked to the previous coefficients $b_i$ by the following set of
relations
\begin{eqnarray}
\label{eq54}
\bar{a}_1&=& - \hbar^2 \frac{b_1}{2 \Delta x_1} \quad , \quad \bar{a}_2 = - \hbar^2
\frac{b_2}{2 \Delta x_2} \quad , \quad  
\bar{a}_3 = - \hbar \langle \hat x_1\rangle \frac{b_1}{ \Delta x_1}  \quad ,
\quad \bar{a}_4 =   - \hbar \langle \hat x_2\rangle \frac{b_2}{ \Delta x_2}
\quad , \nonumber\\   \bar{a}_5 &=& \frac{1}{8} \frac{\theta^2}{\hbar^2}
\frac{b_2}{\Delta x_2} + \frac{b_3}{2 \Delta x_3} \quad , \quad 
\bar{a}_6= \frac{1}{8} \frac{\theta^2}{\hbar^2} \frac{b_1}{\Delta x_1} +
\frac{b_4}{2 \Delta p_2} \quad , \nonumber\\
&\bar{a}_7& = -\frac{\theta}{\hbar} \langle \hat x_2 \rangle \frac{b_2}{2 \Delta
x_2} - \langle \hat p_1 \rangle \frac{b_3}{2 \Delta p_1} \quad , \quad 
\bar{a}_8 = -\frac{\theta}{\hbar} \langle \hat x_1 \rangle \frac{b_1}{2 \Delta
x_1} - \langle \hat p_2 \rangle \frac{b_4}{ \Delta p_2} \quad , \nonumber\\
&\bar{a}_9& = \left( \langle \hat x_1 \rangle^2 - (\Delta x_1)^2  \right)
\frac{b_1}{2 \Delta x_1} + 
\left( \langle \hat x_2 \rangle^2 - (\Delta x_2)^2
 \right) \frac{b_2}{2 \Delta x_2} + 
\left( \langle \hat p_1 \rangle^2 - (\Delta p_1)^2  \right)
\frac{b_3}{2 \Delta p_1}  \nonumber\\
& & +  
\left( \langle \hat p_2 \rangle^2 - (\Delta p_2)^2  \right)
\frac{b_4}{2 \Delta p_2}   \quad .
\end{eqnarray} 

To summarize,  states which are  extrema of the action given in
Eq.(\ref{eq48})  satisfy the  equation
displayed in Eq.(\ref{eq53}). The real quantities appearing in this 
equation are linked to the observables of these state-solutions by the two sets
of relations given in Eqs.(\ref{eq50},\ref{eq54}). As done in the previous 
section, we introduce an operator depending on the state 
$\vert \psi \rangle$ and apply it to a state $\vert \phi \rangle$. This amounts to 
replace $\psi(p)$ by $\phi(p)$ in Eq.(\ref{eq54}) while the coefficients $a_k$ depend
on $\vert \psi \rangle$. This is the equivalent of Eq.(\ref{eq1000}).

Before proceeding, let us point out one important difference between our
approach and the  usual generalization of coherent states. The operators
$\hat x_i, \hat p_j$ of the non commutative plane can be written as linear
combinations of some  $ \hat Q_j, \hat P_j$ which obey the usual commutation relations.
As discussed in \cite{pere}, the concept of coherent states is
defined for the algebra and its representation. This means  it can not
be affected by a change of generators. On the contrary, our approach to what we
call {\sl squeezed} states depends on the
generators, as shown by the mixing of $p_1$ and $p_2$ in Eq.(\ref{eq53}). This is hardly
surprising since the action we began with, although it physically makes sense, is
not invariant for linear transformations on the phase space. One can use the
operators $\hat Q_j, \hat P_j$ to show the impossibility to saturate 
simultaneously the three uncertainties analyzed in section $2$; we simply 
restricted ourselves
to a reasoning which could  also be used for the other models where the link to an 
undeformed algebra is less obvious.

The change of variables
performed in Eq.(\ref{eq46}) is not well defined in the limit $\theta
\longrightarrow 0$ and the action $S$ does not coincide with two copies of the
one dimensional case displayed in Eq.(\ref{eq23})(generalized  coherent states
also become singular when the non parameter controlling non commutativity is sent 
to zero). However, the
differential equation it generates has that nice property. 
Looking at Eq.(\ref{eq53}), one sees that the factors multiplying the terms
$p_1 \partial_{p_2}$ and $ p_2 \partial_{p_1}$ go to zero in that limit so that
 one obtains  the sum of two separate operators ${\cal O}_{x_1,p_1}$ and  
${\cal O}_{x_2,p_2}$ just like in Eq.(\ref{nontri}). 
Using
 Eqs.(\ref{eq50},\ref{eq54}), one sees that the coefficients 
 $\bar a_1, \bar a_2 $ of the derivatives $\partial_{p_1}^2, \partial_{p_2}^2$
 admit expansions in the deformation parameter $\theta$. This means a 
 perturbation theory in this context will not be straightforward.

As in the previous section, we will take  quantities $a_k$ depending on a state 
$\vert \psi \rangle$ and write the associated  differential equation for a state
$\vert \phi \rangle$.
We don't have the most general solution to this second order partial
differential equation. We shall look for
special cases in which explicit solutions can be found. It is rather interesting
that the states found in Eq.(\ref{eq102}-\ref{eq103}) which display  periodic 
behaviors for the harmonic oscillator are particular solutions of the aforementioned equation.

It is obvious that the coefficients $b_i$ are positive. As a consequence, one
has
\be
\label{eq55}
\bar{a}_1 , \bar{a}_2  < 0  \quad , \quad \bar{a}_5 , \bar{a}_6  > 0  \quad .
\ee 
We shall take into account the signs of these coefficients by the following 
parametrization:
\newline  $\bar{a}_1= - a_1^2 , \bar{a}_2= - a_2^2  , 
\bar{a}_5=  a_5^2  , \bar{a}_6=  a_6^2$. To simplify future formulas, 
we  assume from now on $\hbar = \theta = 1.$ Introducing the variables 
$y_1,y_2$  by the relations
\be
\label{eq56}
p_1 = a_1 y_1 - \frac{a_4}{a_2^2} \quad , \quad p_2 = a_2 y_2 + \frac{a_3}{a_1^2} 
\quad , 
\ee
our differential equation takes the simpler form
\be
\label{eq57}
\left[ - \partial^2_{y_1} -  \partial^2_{y_2} +
 i t_1^2 ( - y_2 \partial_{y_1} + y_1 \partial_{y_2}) +
  ( t_2^2  y_1^2 + t_3^2   y_2^2
+ t_4  y_1 + t_5  y_2 + t_6 ) \right] \phi(y) = 0  \quad , 
\ee
where
\begin{eqnarray}
\label{eq58}
t_1^2 &=& a_1 a_2 \quad , \quad t_2^2= a_1^2 a_5^2 \quad , \quad t_3^2= a_2^2 a_6^2 
\quad , \quad t_4 = \left( - 2 \frac{a_5^2}{a_2^2} a_4 + a_7 \right) a_1 \quad ,
\quad  t_5 = \left(  2 \frac{a_6^2}{a_1^2} a_3 + a_8 \right) a_2 \quad , \nonumber\\
t_6&=& a_9 + \left( \frac{a_4 a_5}{a_2^2} \right)^2 -   \frac{a_4 a_7}{a_2^2} +
\left( \frac{a_3 a_6}{a_1^2} \right)^2 +   \frac{a_3 a_8}{a_1^2} \quad .
\end{eqnarray}

\subsubsection{The Gaussian solution}
  
Similarly to the one dimensional case, one can, {\em in this case}, postulate a
solution which is the exponential of a quadratic function:
 \be 
\label{eq59}
\psi(p) = N \exp{( z_1 y_1^2 + z_2 y_2^2 +z_3 y_1 y_2 + z_4 y_1 + z_5 y_2)}
\quad .
 \ee 
Plugging this wave function in the differential equation, one ends up with the relations
\begin{eqnarray}  
\label{eq60}  
z_3 &=& \frac{t_1^2}{4(t_2^2-t_3^2)} 
\left( \sqrt{-(t_1^2 - 4 t_2^2)(t_1^2 - 4 t_3^2)} +
 i ( t_1^4 - 2( t_2^2+t_3^2)) \right)  \quad , 
 \quad z_1 = \frac{1}{2} \sqrt{- z_3^2 - i z_3 t_1^2+ t_2^2} \quad , \nonumber\\ 
z_2 &=&  \frac{1}{2} \sqrt{- z_3^2 - i z_3 t_1^2+ t_3^2} \quad , \quad 
z_5= \frac{2 z_3 t_4 + i t_1^2 t_4 - 4 z_1 t_5}{16 z_1 z_2 - 4 z_3^2 - t_1^4}
 \quad , \quad z_4= - \frac{2 z_3 z_5 - i z_5 t_1^2 - t_4}{4 z_1} \quad , \nonumber\\
 & & 2 z_1 + 2 z_2 + z_4^2 + z_5^2 - t_6 = 0 \quad .
\end{eqnarray} 
Each coefficient has been  written solely in terms of those
appearing before it in the list and the first one, $z_3$, depends only on the
 parameters of the equation. The last formula of Eq.(\ref{eq60}) is a constraint
  which has to be satisfied for the
differential equation   to admit a Gaussian solution; this is reminiscent of 
Eq.(\ref{eq1004}) in the usual one dimensional case. The normalizability of the state  
imposes conditions on  $z_1,z_2,z_3$.

\subsubsection{The cylindrically symmetric solution}

In the preceding subsection we saw that a carefully chosen relation between the
coefficients of the differential equation led to an explicit solution.
Although in the usual one dimensional case  such a choice
was ultimately justified because it led to the absolute minimum of the action, nothing like
that occurs here. One has to resort to a second order analysis to 
find the true
 nature of the critical points represented by the states obtained so far. That will
  not be  done here; we simply find some explicit solutions. 

 There is a set of simple relations between the parameters of the equation
  which allows a cylindrical separation of
  variables. In fact, if 
\be
\label{eq61}
t_3 = t_2 \quad , \quad t_4 = t_5 = 0  \quad  ,
\ee
the parameterization $\phi = R(r) e^{i m \theta} $ in the polar coordinates linked to
the Cartesian coordinates $(y_1,y_2)$ leads to the ordinary differential equation

\be
\label{eq62}
R''(r) + \frac{1}{r} R'(r) + \left( \frac{m^2}{r^2} + (t_6 - m t_1^2) + t_2^2 r^2
  \right) R(r) = 0 \quad .
\ee
In the special case where the relation $t_6 = m t_1^2$ holds, the solution can be 
re casted as a combination of two Bessel functions:
\be
\label{eq63}
R(r) = C_1  I_{- \frac{m}{2}} \left(\frac{t_2 r^2}{2} \right) +
 C_2 I_{ \frac{m}{2}} \left(\frac{t_2 r^2}{2}  \right) \quad .
\ee

\subsubsection{A third explicit solution}
  
  Let us consider the case 
  \be
  \label{eq65}
  (t_1,t_3,t_4,t_5,t_6) = (2,2,0,0,-4) \quad , \quad {\rm with} \quad  t_2 \quad
   {\rm arbitrary} \quad .
  \ee
  It is straightforward to verify that for any integer $m$ and any 
   complex constants $c_{01},c_{11}$, the function
\be
\label{eq66}
\phi(r,\theta) = r^{m} \exp{ \left[ - c_{01} e^{i \theta} r - r^2 \left(1 + 
\frac{1}{4} c_{11} e^{2 i \theta} \right) + i m \theta \right] }
\ee
is a solution of Eq.(\ref{eq57}).

\subsubsection{Factorizability}

 As in the undeformed theory, we can look for the particular conditions  under which the
 second order differential operator can be written as a product. This is found to 
 occur  when

\be
\label{eq67}
t_2 = t_3 = \frac{t_1^2}{2} \quad , \quad t_6 =  t_1^2  \quad , \quad t_4 = t_5 = 0 
\quad .
\ee
The operators $k_1,k_2$ assume the forms
\be
\label{eq68}
k_1 = \left(  i \partial_{y_1} + \partial_{y_2} - 
\frac{i}{2}  t_1^2 (y_1 - i y_2 ) \right)  \quad , 
\quad k_2 = \left(  i \partial_{y_1} - \partial_{y_2} + 
\frac{i}{2}  t_1^2 (y_1 + i y_2 ) \right)  \quad .
\ee
This case of reducibility forms a particular subset which is at the intersection of the
ones corresponding to Gaussian solutions and the ones displaying  polar symmetry; it 
does not bring in new solutions.

\subsection{The model with a minimal uncertainty in length}

We now restrict to the second extension of the KMM model corresponding to the following
choice of the functions appearing in Eq.(\ref{eq12}): 
\be
\label{eq90}
 g(\vec{p}^2) = \beta \quad , \quad  f(\vec{p}^2) = \frac{ \beta \vec{p}^2}{-1+
  \sqrt{1+2 
  \beta \vec{p}^2}} \quad .
\ee
The main interest of this model lies in the fact
 that its Q.F.T is finite \cite{kmm3}. It admits  a representation in which 
 the operators look much simpler
 than in Eq.(\ref{eq15}):
 \be
 \label{91}
 \hat x_i = i \hbar \partial_{\rho_i} \quad , \quad p_i =
  \frac{\rho_i}{1-\frac{1}{2} \beta \vec{\rho}^2} \quad ;
 \ee
 the  scalar product is the usual one, but now it is defined on the disk of radius
 $\sqrt{\frac{2}{\beta}}$. 
The couples which enter the action $S$ are  $(x_1,p_1),(x_1,p_2),(x_2,p_1)$
and $(x_2,p_2)$. The reasoning we have used in the preceding sections lead to the
final equation
\begin{eqnarray}
\label{eq92}
0&= & \left[ \left( 1- \frac{1}{2} \beta \vec{\rho}^2 \right)^2 ( -a_1^2 \partial_{\rho_1}^2- 
a_2^2 \partial_{\rho_2}^2) + \frac{2 i}{\hbar} 
\left( 1- \frac{1}{2} \beta \vec{\rho}^2 \right)
(- a_1^2 \langle \hat x_1\rangle \partial_{\rho_1}  -
 a_2^2 \langle \hat x_2\rangle \partial_{\rho_2}) \right. \nonumber\\
&+& \left.  a_3 \rho_1^2 + a_4 \rho_2^2 + a_5 \rho_1 \rho_2 + 
(a_6 \rho_1+a_7 \rho_2) \left( 1- \frac{1}{2} \beta \vec{\rho}^2 \right)+
a_8 \left( 1- \frac{1}{2} \beta \vec{\rho}^2 \right)^2  \right] 
\psi(\rho_1,\rho_2) \quad .
\end{eqnarray}

The symmetric case $ a_2 = a_1 \quad , \quad \langle x_1 \rangle = 
 \langle x_2 \rangle = 0 \quad , \quad
a_3 = a_4 = a_5 = a_6 = a_7 = 0 $ admits a solution which
is a combination of Bessel functions multiplied by phases:
 \be
\label{eq93}
\psi(\rho) = \left( C_1 I_{-m} \left(\frac{\sqrt{- a_8}}{a_1} \rho \right)+ 
 C_2 I_{m} \left(\frac{\sqrt{- a_8}}{a_1} \rho \right) \right) e^{i m \theta} \quad .
\ee 
The coefficients $a_6,a_7$ and $a_8$ vanish with $\beta$.

\section{A look at the fuzzy sphere}

The fuzzy sphere is a matrix model defined by  the following relations 
\cite{coh4}:
\be
\label{eq109}
[ \hat x_k , \hat x_l ] = \frac{i \alpha}{\sqrt{j(j+1)}} \epsilon_{klm} \hat x_m \quad ; \quad 
\hat x_1^2 + \hat x_2^2 + \hat x_3^2 = 1  \quad , \quad
{\rm with} \quad j \quad  {\rm half-integer} \quad .
\ee 
The saturation of the uncertainties related to the pairs of non commuting variables
translate into the formulas  $ m_{jk} \vert \psi \rangle = 0 $, with
\be
\label{eq110}
m_{12} = \hat x_1 + i a \hat x_2 + (d+ i e ) \quad , \quad
m_{23} = \hat x_2 + i b \hat x_3 + (f+ g e ) \quad , \ {\rm and} \quad
m_{31} = \hat x_3 + i c \hat x_1 + (h+ i k ) \quad ,
\ee
where $a,b,c, \cdots$ are real. Considering the following combinations of these equations
\be
\label{eq111}
(c [ m_{12}, m_{23} ] - i  [ m_{23}, m_{31} ]) \vert \psi \rangle = 0 \, , \, 
(- i [ m_{12}, m_{23} ] + b [ m_{31}, m_{12} ]) \vert \psi \rangle = 0 \, , \,
(- a [ m_{23}, m_{31} ] + i  [ m_{31}, m_{12} ]) \vert \psi \rangle = 0 \, , \,  
\ee
one obtains( in units where $\alpha=1$) 
\be
\label{eq112}
(1-i a b c) \hat x_1 \vert \psi \rangle = 0 \quad , \quad
(1-i a b c) \hat x_3 \vert \psi \rangle = 0 \quad , \quad
(1-i a b c) \hat x_2 \vert \psi \rangle = 0 \quad .
\ee
Can the  three Heisenberg inequalities  be saturated  simultaneously? Only two cases
may lead to that situation: 
\begin{itemize}
 \item The first possibility is
        \be
        \label{eq113}
          \hat x_1 \vert \psi \rangle = \hat x_2 \vert \psi \rangle = 
          \hat x_3 \vert \psi \rangle = 0 \quad ,
        \ee
     but then the second part of Eq.(\ref{eq109}) is violated \quad .
 \item The remaining possibility 
       \be
       \label{eq114}
          1 - i a b c = 0 
       \ee 
 can be rewritten as  
 \be
 \label{eq115}
 \frac{ \langle \hat x_1 \rangle \langle \hat x_2 \rangle  \langle \hat x_3 \rangle}
 {(\Delta x_1)^2 (\Delta x_2)^2 (\Delta x_3)^2 } = - 8 i \quad ;
 \ee
 this is not possible since
 all the quantities on the left side  are real.
\end{itemize} 

  The method proposed here can in principle be applied to  the fuzzy
 sphere. In this section, we work  with an action slightly different from  the 
 ones used 
 so far, choosing $m=1$:
 \be
 \label{eq116}
 S = \left[ \left( (\Delta x_1)^2 (\Delta x_2)^2 - \frac{1}{2} \langle x_3 \rangle^2 \right)^2 +
  {\rm permutations} \right] 
   \quad .
 \ee
 One important feature which distinguishes the fuzzy sphere from the 
 models we have studied before is the fact that its Fock space is finite dimensional. This 
 results in the fact that the action $S$ is now a function rather than a functional.
 
 To illustrate how our approach applies to the fuzzy sphere, let us take the very 
 simple case $j=1$. A unitary transformation is performed to go from the usual 
 representation to a more symmetric 
 one in which the non vanishing elements of the operators are
 \be
 \label{eq117}
 (X_1)_{23} = (X_1)_{32} = 1 \quad , \quad  (X_2)_{13} = - (X_2)_{31} = -i \quad , \quad
 (X_3)_{12} = (X_3)_{21} = 1 \quad .
 \ee
 The Fock space in this trivial case is six dimensional on the real scalars.
 Any state can be written as
 \be
 \label{eq118}
   \vert \psi \rangle = ( z_1, z_2 ,  z_3 , z_4, 
  z_5, z_6 ) \quad .
  \ee  
 The mean values we need in order to compute the action can be written as
 \be
 \label{eq119}
 \langle x_k \rangle = \frac{N_k}{G} \quad , \quad 
  \langle x_k^2 \rangle = \frac{K_k}{G} \quad ,
 \ee
 where
 \begin{eqnarray}
 \label{eq120}
 N_1 &=& 2 (z_3 z_5 + z_4 z_6) \quad , \quad N_2 = 2 (- z_2 z_5 + z_1 z_6 ) \quad , 
 \quad N_3 = 2 ( z_1 z_3 + z_2 z_4 ) \quad , \nonumber\\
 K_1 &=& z_3^2 + z_4^2 + z_5^2 + z_6^2 \quad , 
 \quad K_2 = z_1^2 + z_2^2 + z_5^2 + z_6^2
  \quad , \quad K_3 = z_1^2 + z_2^2 + z_3^2 + z_4^2 \quad , \nonumber \\
 G &=& z_1^2 + z_2^2 + z_3^2 + z_4^2 + z_5^2 + z_6^2 \quad . 
 \end{eqnarray}
 The interest of the representation displayed by Eq.(\ref{eq117}) is that it makes the 
 symmetries of the problem more transparent. One easily verifies that the 
 transformation 
 \be
 \label{eq121}
 \tau( z_1, z_2 , z_3, z_4, 
  z_5, z_6 ) = ( z_3 , z_4 , z_5 , z_6 , -z_2,z_1) 
 \ee
generates a circular permutation(discarding the signs) of the functions $N_k$. The same
 holds for the functions $K_k$ while $G$ remains unchanged so that $\tau$ is a 
 symmetry of the action and, in addition, it is idempotent: $\tau^6=- 1$.
 Finding the extrema of the action given above is not an easy task. 
 We give here two such extrema, obtained numerically using Mathematica:
 \begin{eqnarray}
 \label{eq122}
 \vec z &=& (-1,0,0,1,1,0) \quad {\rm with } \quad  S = \frac{16}{27} \quad {\rm and} \nonumber\\ 
 \vec z &=& (0.631646,0.315353,0.002528,0.016494,-0.316359,0.633415)
  \quad  {\rm with} 
\quad  S = 1.19 \, 10^{-8} \quad .
 \end{eqnarray}
 Applying the transformation $\tau$ to these states, one generates states  having the same 
 values of $S$.
\section{conclusions}

In this work, we have shown that the defining properties of the coherent
states in the usual theory are lost in some non commutative models. This led
us to suggest an approach toward  squeezed states which
relies on a functional.
We have found special solutions to the second order differential equations obtained in two
 different 
non commutative theories. We have briefly outlined how our method can be applied to
 finite dimensional matrix models like the fuzzy sphere. 
 
 The problem for the first two
 theories we have studied is that there are too many solutions to the relevant
  equations. On the contrary, for the fuzzy sphere, the situation
   is different; the action
is a function of a finite number of variables but its form is non trivial and 
makes the search for solutions cumbersome.
  
One of the crucial points which remain to be addressed is the nature of the
critical points found here. To know if these states  are maxima, minima or saddle
points of the
 action, one has to resort to a second order analysis. However, as the most general
solutions of the second order partial differential equations involved are not known, such
a computation cannot tell us by itself if we are in front of an absolute minimum. One
can also develop the differential equations we obtained order by order 
in the extra parameter $\theta $; this breaks the symmetry with $\hbar$ and makes any
statement about the nature of the solution  more difficult. One important
difference with usual perturbation theory lies in the fact that the
deformation parameter appears also multiplied by derivative operators. This is
likely to be tackled by a treatment such as the one used in \cite{muso}.

Some questions are of particular importance for the approach we
suggest to be really valuable. For example, one would like to know if
the states we obtained  form an over complete system. If If this was the
 case, they might be legitimate
candidates for the definition of a physically meaningful star product 
\cite{voros}. The fact that some solutions obtained here 
 are eigenfunctions of Sturm-Liouville systems is  
promising.

\underline{Acknowledgments}
 
It is a pleasure to thank S.Detournay, Cl.Gabriel and Ph.Spindel for useful
 discussions in Mons(Belgium) in the early stages of this work.

\end{document}